# Electric field modulated topological magnetoelectric effect in Bi₂Se₃

Mintu Mondal,[1, 2, *] Dipanjan Chaudhuri,[1] Maryam Salehi,[3] Cheng Wan,[1, 4]
N. J. Laurita,[1] Bing Cheng,[1] Andreas V. Stier,[1] Michael A. Quintero,[1] Jisoo Moon,[5]
Deepti Jain,[5] Pavel P. Shibayev,[5] Jamie Neilson,[1, 4] Seongshik Oh,[5] and N. P. Armitage[1]

[1] *The Institute for Quantum Matter, Department of Physics and
Astronomy, The Johns Hopkins University, Baltimore, MD 21218, USA*
[2] *School of Physical Science, Indian Association for the Cultivation of Science, Jadavpur, Kolkata 700032, India*
[3] *Department of Material Science and Engineering, Rutgers University of New Jersey, Pscataway, NJ 08854*
[4] *Department of Chemistry, Johns Hopkins University, Baltimore, Maryland, 21218, USA*
[5] *Department of Physics and Astronomy, Rutgers University of New Jersey, Piscataway, NJ 08854*

Topological insulators have been predicted to exhibit a variety of interesting phenomena including a quantized magnetoelectric response and novel spintronics effects due to spin textures on their surfaces. However, experimental observation of these phenomena has proved difficult due to the finite bulk carrier density which may overwhelm the intrinsic topological responses that are expressed at the surface. Here, we demonstrate a novel ionic gel gating technique to tune the chemical potential of Bi₂Se₃ thin films while simultaneously performing THz spectroscopy. We can tune the carrier concentration by an order of magnitude and shift the Fermi energy, E_F to as low as ≃ 10 meV above the Dirac point. At high bias voltage and magnetic field, we observe a quantized Faraday angle consistent with the topological magnetoelectric effect that can be tuned by ionic gel gating through a number of plateau states.

Three dimensional topological insulators (TIs) are characterized in the ideal case by gapped insulating bulk and gapless conducting surface states that have a Dirac-like energy-momentum dispersion relation [1–4]. These topological surface states (TSSs) are expected to exhibit novel topological phenomena including quantized transport and manifest a quantized axion magnetoelectric effect [4–10]. While sample growth has improved such that quantized responses of TSSs have been measured, relatively large charge densities remains a pressing issue in many systems [9, 11, 12]. Therefore, there is a need for sample growth independent method of reducing the bulk conduction.

Magnetoelectrics are materials in which a polarization can be created by an applied magnetic field or a magnetization can be created by an applied electric field [13] and have been topics of research for decades. We recently observed the quantized topological magnetoelectric effect (TME) in thin films of Bi₂Se₃ [9]. The TME rests on a formulation where the electrodynamics of a TI is regarded as a bulk magnetoelectric and not as a surface conductor per se. This can be seen in the following fashion. When a field breaks the time reversal symmetry, the TI surface states will become gapped and allow a surface quantum Hall effect. For a pure Hall current, an applied electric field in the $\hat{y}$ direction will induce a transverse Hall surface current $K_x$. As a surface current can be written as bulk magnetization e.g. $\mathbf{K} = \mathbf{M} \times \hat{n}$ (where $\hat{n}$ is a surface normal), an applied electric field will induce a current that is equivalent to a bulk magnetization. For an inversion symmetric TI, one can show that the surface Hall conductance must be $G_{xy} = (N + \frac{1}{2})\frac{e^2}{h}$ and hence the proportionality between the polarization and applied magnetic field or the magnetization and applied electric field is quantized in terms of universal constants of nature [14] e.g. $M_y = (N + \frac{1}{2})\frac{e^2}{h}E_y$. In this expression $N$ corresponds to the number of occupied gapped states (Landau levels) on the surface due to finite surface electron density. In the TME formulation these can be seen as some number of Chern insulator layers absorbed to the surface. However the $\frac{1}{2}$ arises purely from the bulk magnetoelectric effect as any strictly two dimensional insulator must have a quantum Hall conductance that is $\frac{e^2}{h}$ times an integer. Hence the surface states must not be seen as a 2D system, but as a boundary of a three-dimensional solid. This is the quantized response demonstrative of TIs being a distinct state of matter.

The TME can be measured as quantized Faraday and Kerr rotations when time-reversal symmetry is broken. In the present work we demonstrate that we can tune the surface state contribution to the TME. We do this by developing a novel ionic gel gating technique through which we can tune the Fermi level very close to the Dirac point, while studying the magnetoelectric effects using time domain THz spectroscopy (TDTS). Unlike more conventional metallic gates, the ionic gel gating is effectively transparent to THz radiation at low temperatures and fully compatible with TDTS. As a function of magnetic field, we observe both semiclassical and quantized transport regimes. The low field semiclassical cyclotron resonances provide a way of quantifying the shift of the chemical potential on the surface. For low carrier densities and at high enough magnetic fields, we observe a quantized Faraday rotation consistent with the TME. The TME can be tuned by applied electric field from the ionic layer.

High quality Bi₂Se₃ thin films were prepared using molecular beam epitaxy as described by Koirala *et al.*

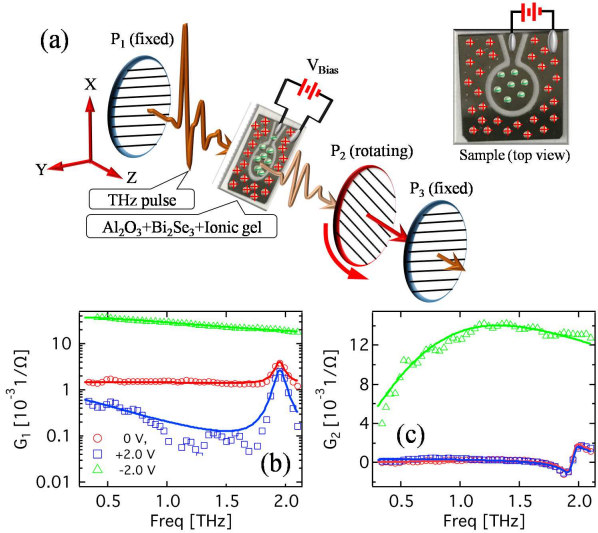

FIG. 1. (a) Schematic of the experimental setup to study the TME of the topological surface states tuned by electric field. $P_1$ (fixed), $P_2$(rotating), and $P_3$ (fixed) are the wire grid polarizers for measuring the $\theta_F$ of the THz pulse. (b-c) Real and imaginary parts of the conductance ($G = G_1 + iG_2$) for 15 QL Bi$_2$Se$_3$/MoO$_3$ film measured at different bias voltages.

[11]. Ionic liquid gating is one of the most effective ways to apply a high electric field to tune the Fermi level [15, 16]. Additionally, the method has the particular advantage for the present case that the conducting electrodes are out of the THz beam path. Moreover, at low temperatures a thin ionic liquid layer is largely transparent at THz frequencies. The high polarizability of the ionic liquid ensures a uniform electric field at the surface as shown in Fig. **??** of the supplementary information (SI). However, liquid gates cannot be used in vertical optical setups. The liquid gates can also have nonuniform thicknesses (droplet) and variable optical shape upon temperature cycling thereby variable optical properties. This renders typical liquid gating approaches incompatible with TDTS spectroscopy. To overcome these problems, we have developed a novel method to prepare an ionic *gel* that can be spin coated on the TI film and allowed to dry. The ionic liquid is dissolved in a co-polymer that ensures mechanical stability without impeding the capacity to apply large E-fields. To apply a gate bias, films were patterned by photolithography and wet etching in diluted aqua regia solution, HNO$_3$/HCL/H$_2$O (1:3:2). Each patterned film consists of a circular region which is used for THz transmission spectroscopy and the surrounding region which is used as a gate electrode. See Fig. 1 and the SI for further details of the configuration. The bias was applied at 280 K and then cooled down, effectively freezing in the applied electric field.

TDTS allows the measurement of both components of the optical conductance over a range of approximately 0.1 - 2.5 THz and and also the sample induced Farday and Kerr rotation of THz radiation with high precision [17]. We measured the complex conductance of Bi$_2$Se$_3$ thin films at different bias voltages. In Fig. 1 (b-c) we show the real and imaginary part of the conductance of a 15 quintuple layer (QL, 1 QL $\approx$ 1 nm) Bi$_2$Se$_3$ film capped with a 20 nm thick amorphous MoO$_3$ layer, measured at 5 K with different applied voltages. MoO$_3$ has been previously shown to reduce the charge density of Bi$_2$Se$_3$ films due to its charge accepting properties [18, 19]. Here presumably it does that and it also protects the Bi$_2$Se$_3$ sample against chemical reaction with the ionic gel that we found occurs without it at even moderate voltages (See SI). Although the protective layer slightly reduces the electric field strength, it also allows us to increase the bias voltage to higher values and provides better control.

At zero voltage, the conductance is almost unchanged as compared to the bare film without any ionic gel [9, 20]. It is characterized by a low frequency Drude peak and a prominent phonon $\simeq$1.95 THz. As shown in Fig. 1 (b-c), for applied positive bias, $V_{bias} = +2.0$ V, the conductance of the sample is reduced to a very low value which indicates carrier depletion, whereas for a negative bias of -2.0 V, we observed an almost 50 fold increase of conductance. The sample in this configuration shows a strong Drude peak at low frequency, whereas the phonon contribution gets screened by strong Drude response. This is consistent with our interpretation that a negative bias voltage moves the chemical potential upwards into the conduction band, thereby increasing the conductance.

We can fit these conductance curves with a three component Drude-Lorentz model (see details in the SI) and obtain excellent fits (solid lines in Fig. 1 (b-c)) to the data. According to the spectral weight sum rule (see SI) the Drude spectral weight is related to the carrier density via $\omega_{pD}^2 d = \frac{n_{2D}e^2}{m^*\epsilon_0}$ . Using this formula, one can estimate a two dimensional carrier density of $n_{2D} \simeq 7.8 \times 10^{12}$ cm$^{-2}$ corresponding to the Fermi energy, $E_F \simeq$213 meV above the Dirac point at zero bias voltage. Here $m^*$ is the effective mass of the TSS fermions that is given by $\frac{\hbar k_F}{v_F}$, which can be estimated by the TSS dispersion relation $E_k = Ak_F + Bk_F^2$ (with quadratic corrections) where $A$ and $B$ are parameters estimated from ARPES [21] and given in [22]. With changing $V_{bias}$ to +2.0 V, $n_{2D}$ is depleted to $2.1 \times 10^{11}$ cm$^{-2}$ and an $E_F \simeq$24 meV above the Dirac point. With applied −2.0V the large increase in conductance is consistent with a chemical potential shift that is more than enough to put it in the conduction band [19] and hence a simple one band model cannot be used to estimate the carrier density.

After establishing the basic systematics of the ionic gel technique, we now apply it to study in detail the magneto-optics of a similar 8 QL Bi$_2$Se$_3$ sample with 20 nm thick MoO$_3$ capping layer. Fig. 2 shows a sequence of real and imaginary part of $\theta_F$ measured using polariza-tion modulated TDTS technique [19, 23] and a number of different regimes have been observed depending on applied magnetic fields and bias voltages.

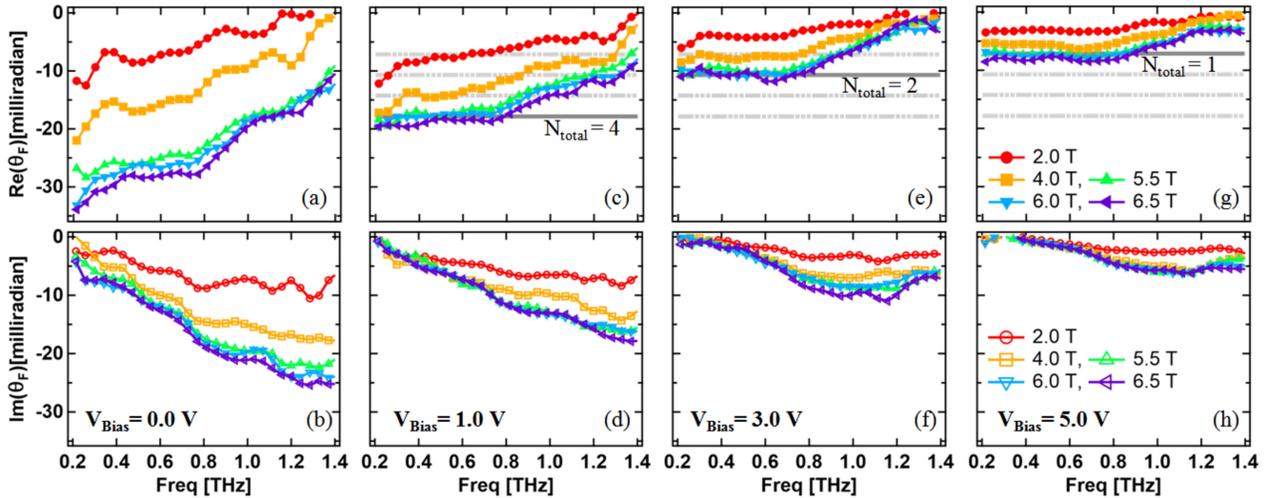

FIG. 2. Real and imaginary parts of the Faraday angle ($\theta_F$) of a 8 QL $Bi_2Se_3/MoO_3$ sample measured at different bias voltages ($V_{Bias}$) and magnetic fields. The solid gray lines represent the real part of the $\theta_F$ expected from Eq. 3 assuming a certain number of filled Landau levels.

At low magnetic fields and/or low bias voltages, one is in a regime of semi-classical transport. A cyclotron resonance feature consistent with reasonably strong damping appears in the imaginary (dissipative) part of the Faraday rotation which then generally moves to higher frequencies with higher magnetic field. One can see ( e.g. red curves) that at a fixed magnetic field, Re[$\theta_F$] decreases with increasing bias voltage which is consistent with decreasing carrier density. In this low bias voltage and magnetic field range, $\theta_F$ is frequency and magnetic field dependent and the thin film's magneto-optical response can be describe by the semi-classical Drude cyclotron resonance model. At low fields, we fit the Faraday angle to the semi-classical Drude expressions in the presence of a magnetic field (see SI for details). Figs. 3 (a-d) show representative fittings for the real and imaginary parts of the Faraday angle. Through these fits we can extract the Drude plasma frequency ($\omega_{pD}$) and the cyclotron frequency ($\omega_c$). In Fig. 3(e) we plot $\omega_c$ as a function of $B$ for different applied voltages.

In the semiclassical transport regime the cyclotron resonant frequency $\omega_c$ is given by $eB/m^*$ with the cyclotron mass given by the expression $m^* = \frac{\hbar^2}{2\pi}\frac{dA}{dE}$ where $A$ is the momentum-space area enclosed by the cyclotron orbit. For the lowest fields and/or high $E_F$, the cyclotron frequency, can be fit to the usual semiclassical expression $\frac{eB}{m^*}$ where $m^*$ can be defined as $\frac{\hbar k_F}{v_F}$ where the Fermi velocity, $v_F = 3.2 \times 10^5$ $m/s$. However for even moderate fields at low carrier density the linear dispersion of the surface state starts to play a role. A notable curvature in

$\omega_c$ vs. magnetic field, which is characteristic of massless Dirac fermions in the semiclassical transport regime is observed. Then, the cyclotron frequency crosses over to a dependence that can be estimated by the difference in energy between the uppermost filled level and the next unoccupied Landau level i.e. by $\hbar\omega_c = E_{N+1} - E_N$. Here, $E_N = \text{sign}(N)v_F\sqrt{2e\hbar B}\sqrt{N}$ stands for the energy of the $N$th Landau level. By substituting $E_F$ for $E_N$, one gets a relation for the cyclotron resonance frequency, which is

$$\hbar\omega_c = \sqrt{(2e\hbar v_F^2)B + E_F^2} - E_F \qquad (1)$$

We fit the measured cyclotron frequency, $\omega_c/2\pi$ as a function of $B$ using Eq.1 with $E_F$ as a fitting parameter, in the regime of low magnetic fields, where the quantized response (discussed below) is not noticeable. The resultant values for $E_F$ are shown in green in Fig. 3 (f). These estimates for $E_F$ can be compared to estimates of the Fermi energy from the Drude spectral weight ($\propto \omega_{pD}^2$) at each bias voltage (see SI). One can see that there is an excellent agreement between these two estimates of the Fermi energy (Fig. 3 (f)) calculated independently. Note that the estimates from the spectral weight can be made either with the full fitted dispersion or with a linear approximation near the Dirac point, but that the differences are negligible.

These measurements show that the Fermi energy can be reduced dramatically with applied bias voltage, $V_{Bias}$, to as low as 10 meV above the Dirac point which is unprecedented. With a finite bias the ionic gel polarizes to form effectively two capacitors in series, each with spacing of the order of the molecular diameter $d$ with areas equal to the area exposed to the ionic gel. The effective chemical potential change of the TI surface states can be accounted for by an additional *quantum capacitance* in series with the two geometric capacitances. The total applied voltage is

$$V_{Bias} = \frac{Qd}{\epsilon_0 \epsilon_s A_{el}} + \frac{Qd}{\epsilon_0 \epsilon_s A_{gate}} + \frac{Q}{e^2 D(E_F)} \quad (2)$$

where $A_{el}$ and $A_{gate}$ are the surface areas of the electrode and gated area of the TI films and $D(E_F)$ is the density of states at the (voltage dependent) Fermi energy. The final term in this expression arises from the quantum capacitance and is the chemical potential change due to applied bias $V_{Bias}$. One can see that only a small fraction of the applied voltage is available to change the potential of the TI layer.

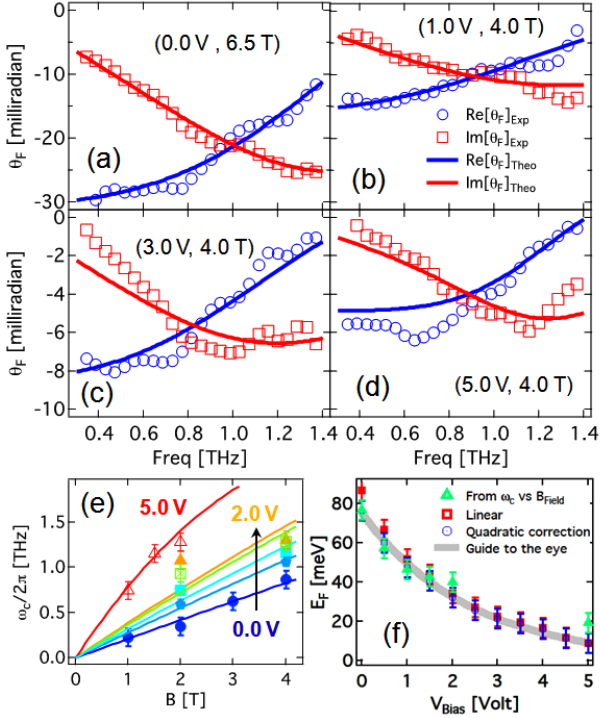

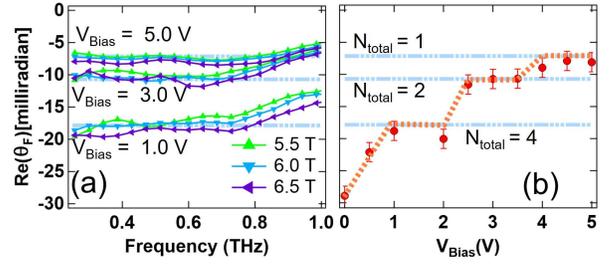

FIG. 3. (a-d) Complex Faraday rotation and fits (as described in the SI) for a 8 QL Bi$_2$Se$_3$/MoO$_3$ sample for different bias voltage. (e) Cyclotron frequency vs $B$ at different bias voltage, $V_{Bias} = (0, 0.5, 1.0, 1.5, 2.0, 5.0)$ V. Fits are using Eq. 1 with the Fermi energy as a free parameter. (f) The Fermi energy as a function of applied voltage, calculated using both the fits from Eq. 1 for the cyclotron frequency as well as from the estimated spectral weight from fitting of complex $\theta_F$ data using both quadratic and linear parametrization to the dispersion.

FIG. 4. (Color online) (a) Real part of Faraday rotation ($\theta_F$) at high magnetic field, $B \geq 5.5T$. The grey lines are theoretically predicted values assuming particular filling factors of the surface states. (b) Average value of Re $[\theta_F]$ over frequency range spanning from 0.2 to 0.8 THz at 6.5 T at different values of the bias voltage ($V_{Bias}$).

We now concentrate on our Faraday angle data at high magnetic fields. In Fig. 2(b-d), we can see that at high magnetic fields, the Faraday angle at low frequency deviates from the smooth cyclotron resonance form and becomes frequency and magnetic field independent. At these fields, the cyclotron resonance has been pushed to higher frequencies, beyond the measured spectral range leaving behind a quantized response. We replot the quantized data in Fig. 4(a). As shown previously [9] the Faraday angle in the quantized regime can be expressed as

$$\text{Re}[\theta_F] \simeq \frac{2\alpha}{n_s + 1} \left( N_b + \frac{1}{2} + N_t + \frac{1}{2} \right) \quad (3)$$

Here $n_s$ is the refractive index of the sapphire substrate ($n_s = 3.09$) and $\alpha$ is the fine structure constant, $\alpha \sim 1/137.04$. The $1/2$ terms arise from the predicted half-quantized contribution to the quantum Hall effects of the top and bottom surfaces. The grey lines in Fig. 4 (a) and (b) are the theoretically predicted values for the quantized Faraday angle from Eq. 3 assuming particular integer values of the Landau level filling factors, $N_{total} = N_b + N_t$. This quantized behavior is that identified earlier as due to the quantized axion magnetoelectric response of a TI or alternatively a quantized Hall conductance of the TI surfaces [9]. In Fig. 4 (b) we plot the average Faraday angle over a frequency range of 0.2 to 0.8 THz as function of bias voltage at 6.5 T magnetic field and a number of distinct plateaus are observed. Consistent with the overall behavior of the Faraday angle with field, the Faraday angle gets smaller with higher bias voltages, reflecting smaller filling factors for lower densities. Generally speaking, at each electric field bias we only see quantized responses at the highest applied magnetic fields (5.5 - 6.5 T); the laboratory magnetic field is not large enough to make a transition to the next quantized plateau.

In this work, we have shown that ionic gel gating can be used to suppresses the bulk carrier density such that the sample enters in the quantized response regime. Using our technique, we can tune the carrier concentration by an order of magnitude and shift the Fermi energy, $E_F$ to as low as $\simeq 10$ meV above the Dirac point. At high bias voltages and magnetic fields, we observe a quantized Faraday angles consistent with the topological magneto-electric effect. It can be tuned by ionic gel gating through a number of plateau states. As discussed above, in Refs. [9, 14] and earlier in Refs. [5, 7], in a system that has inversion in the bulk (as does $Bi_2Se_3$) the magnetoelectric response can be changed by integer multiples of $\frac{e^2}{h}$, without changing anything necessarily about the bulk TME. Physically this corresponds to adsorbing some integer number of Chern insulators ($N_b$ and $N_t$) on the surface or equivalently filling Landau levels of the surface states as expressed in Eq. 3. In our experiment this manifests as the quantization of the Faraday effect as in shown in Fig. 4 (b) which demonstrates that at high fields, the Faraday rotations are quantized in discrete units. These discrete units correspond to the fact that at high fields, surfaces layers can change their Hall conductances only by an amount equal to integer multiples of $\frac{e^2}{h}$.

This work was supported by the NSF EFRI 2-DARE program Grant No. 1542798. Additional support at JHU came from Packard Foundation. We would also like to acknowledge Tyrel McQueen and Liang Wu for helpful conversations.

---

* sspmm4@iacs.res.in

# Supplemental Materials: Electric field modulated topological magnetoelectric effect in Bi₂Se₃

Mintu Mondal,[1, 2, *] Dipanjan Chaudhuri,[1] Maryam Salehi,[3] Cheng Wan,[1, 4]
N. J. Laurita,[1] Bing Cheng,[1] Andreas V. Stier,[1] Michael A. Quintero,[1] Jisoo Moon,[5]
Deepti Jain,[5] Pavel P. Shibayev,[5] Jamie Neilson,[1, 4] Seongshik Oh,[5] and N. P. Armitage[1]

[1] *The Institute for Quantum Matter, Department of Physics and
Astronomy, The Johns Hopkins University, Baltimore, MD 21218, USA*
[2] *School of Physical Science, Indian Association for the Cultivation of Science, Jadavpur, Kolkata 700032, India*
[3] *Department of Material Science and Engineering, Rutgers University of New Jersey, Pscataway, NJ 08854*
[4] *Department of Chemistry, Johns Hopkins University, Baltimore, Maryland, 21218, USA*
[5] *Department of Physics and Astronomy, Rutgers University of New Jersey, Piscataway, NJ 08854*

**Preparation of Bi₂Se₃ samples**

*Photolithography and wet etching*

The Bi₂Se₃ samples used in this experiment are grown using moleculcular beam epitaxy on sapphire subtrates. For details about sample growth, see ref. Koirala *et al.* [S1]. After film growth 20 nm MoO₃ capping was grown to protect against chemical reaction and ageing. Samples were first patterned using standard photolithography processes followed by wet etching. We used photoresist AZ5214E for photolithography and diluted aqua regia HNO₃/HCl/H₂O (1:3:2) to etch our films. Fig. S1(a) shows the general measurement configuration and Fig. S1(b) shows the typical structure of the films after patterning. The circular region in the middle is used for measurement and the surrounding area is used as gate.

*Preparation of ionic gel*

1-ethyl-3-methylimidazolium bis (trifluoromethyl-sulphonyl) imide (C8H11F6N3O4S2, ionic liquid, > 97% by NMR) and polystyrene-b-poly(ethylene oxide)-b-polystyrene (triblock copolymer) were supplied by Sigma-Aldrich and Polymer Source Inc respectively. 350 $\mu l$ ionic liquid was added into a round-bottom Schlenk flask under Argon atmosphere. To remove water from ionic liquid, the Schlenk flask was heated using an oil bath (110 °C) in a dynamic vacuum for 3 days. Next, 40 mg triblock copolymer was added in the flask in the Ar-filled glovebox. At the last step, 2.3 ml anhydrous dichloromethane (VWR, 99.7%) solvent was poured into the flask to dissolve the ionic liquid and polymer. All procedures are performed in an Ar-filled glovebox or using Schlenk techniques to exclude oxygen and water.

*Coating Bi₂Se₃ samples with ionic liquid gel*

Bi₂Se₃ samples were spin-coated with as-prepared ionic liquid gel at 4000 rpm for 30 s in an Ar-filled glove-box. The procedure was repeated for three times to ensure the films were uniformly coated. Then the samples were dried at 40 °C for 5 min in the Ar-filled glovebox.

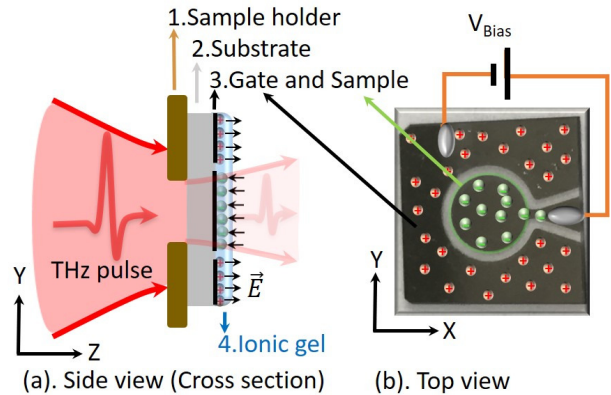

FIG. S1. (Color online) Schematic of the sample. (a) Side view (cross section) shows the transmitted THz pulse passing through the circular region of the sample. The diameter of the aperture of the sample holder is 4.0 $mm$ and the diameter of the circular region of the sample $\simeq 4.1$ $mm$. (b) Top view of the sample shows the surrounding gating region with its electrical connections.

**Fitting conductance data at zero magnetic field**

As shown in Ref. [S2], the conductance data at zero magnetic field can be well fit by three component oscillator model given by the equation,

$$G = \left( \frac{\omega_{\mathrm{pD}}^2}{\Gamma_D - i\omega} + \frac{i\omega\omega_{\mathrm{pDL}}^2}{\omega^2 + i\omega\Gamma_{\mathrm{DL}} - \omega_{\mathrm{DL}}^2} - i\omega\left(\epsilon_{\mathrm{inf}} - 1\right) \right) d\,\epsilon_0 \tag{S1}$$

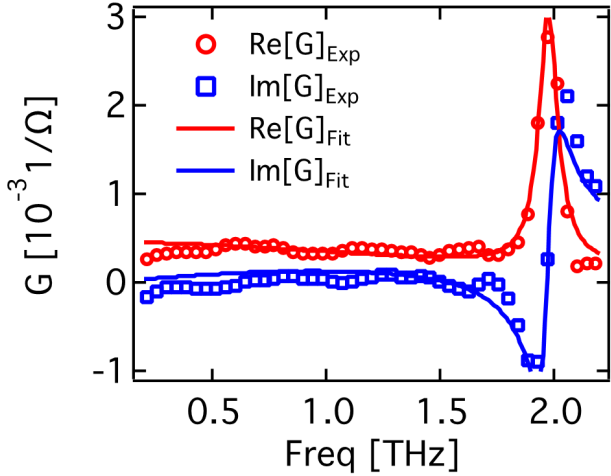

FIG. S2. (Color online) Real and imaginary part of conductance ($G = G_1 + i\,G_2$) of the 8 QL Bi$_2$Se$_3$/MoO$_3$ sample at 5 K measured at zero bias voltage. Solid lines are the fits to the model of Eq. S1.

The first term represents the Drude part describing free-electron like motion and second term is a Drude-Lorentz term that describes the low frequency phonon mode. The last term accounts for the low frequency polarizability, which arise from absorptions above the measured spectral range. Here $\omega_p's$ are the plasma frequencies, $\Gamma$'s are scattering rates, and $d$ is the film thickness. Using the sum rule one can estimate the carrier density from the Drude spectral weight ($\omega_p^2 d$) using the sum rule,

$$\frac{2}{\pi\epsilon_0} \int G_{\mathrm{D1}}\,\mathrm{d}\omega = \omega_{\mathrm{pD}}^2 d = \frac{n_{2D}e^2}{m^*\epsilon_0} \tag{S2}$$

Here $m^*$ is the effective mass of the TSS fermions given by $\frac{\hbar k_F}{v_F}$. To estimate $m^*$ one can use the TSS dispersion relation $E_k = Ak_F + Bk_F^2$. (e.g. the TSS dispersion with quadratic corrections) where $A$ and $B$ are parameters estimated from ARPES as 2.02 eV·Å and 10.11 eV·Å$^2$ respectively whereas $E_k$ and $k_F$ are in eV and Å$^{-1}$ [S3, S4].

The Fig. S2 shows the conductance of the 8 QL sample at zero magnetic field and corresponding fitting to the experimental data using the above Eq. S1. From fitting of the conductance data, the estimated value of the plasma frequency, $\omega_{pD}$ = 43 THz and scattering rate, $\Gamma_D = 1.79$ THz for the surface electrons. For the Drude-Lorentz component and the higher order modes, the fitting gives $\omega_{pDL}$ = 25.3 THz, $\omega_{DL}$ = 1.97 THz, $\Gamma_{DL} = 0.1$ THz and $\epsilon_{\mathrm{inf}} = 80$.

**Fitting Faraday rotation data**

The Faraday rotation is fit by the following expression

$$\tan\left(\theta_{\mathrm{F}}\right) = -\frac{i\left(t_+ - t_-\right)}{t_- + t_+}. \tag{S3}$$

Here $t_\pm$ is the transmission for left and right circularly polarized light. As left and right circularly polarized light are the eigenpolarizations for transmission through the films in magnetic field, their transmission coefficients are given by

$$t_\pm = \frac{(1 + n_s)\exp\left(-i\phi_s\right)}{(1 + n_s) + Z_0 G_\pm} \tag{S4}$$

Here $e^{-i\phi_s}$ is the phase accumulated inside the substrate. $G_\pm$ is conductivity of the film in circular basis in magnetic field that is given by

$$G_\pm = \left( \frac{\omega_{\mathrm{pD}}^2}{\Gamma_D - i\omega \pm i\omega_c} + \frac{i\omega\omega_{\mathrm{pDL}}^2}{\omega^2 + i\omega\Gamma_{\mathrm{DL}} - \omega_{\mathrm{DL}}^2} - i\omega\left(\epsilon_{\mathrm{inf}} - 1\right) \right) d\epsilon_0 \tag{S5}$$

Using the plasma frequencies, phonon frequency, and $\Gamma$s determined at zero field we can fit the Faraday rotation to determine the cyclotron frequency $\omega_c$ which is equal to $\frac{eB}{m^*}$. This allows a model independent measure of $m^*$.

**Degradation of films at high bias voltages**

Induced chemical reactions of the ionic liquid with the film surface at high bias voltage is one of ongoing problems of using ionic liquid gating [S5]. At high bias voltage the ionic liquid reacts with Bi$_2$Se$_3$ films. To prevent the chemical reaction, we used MoO$_3$ protection layer and we have managed to apply more than 5 V routinely.

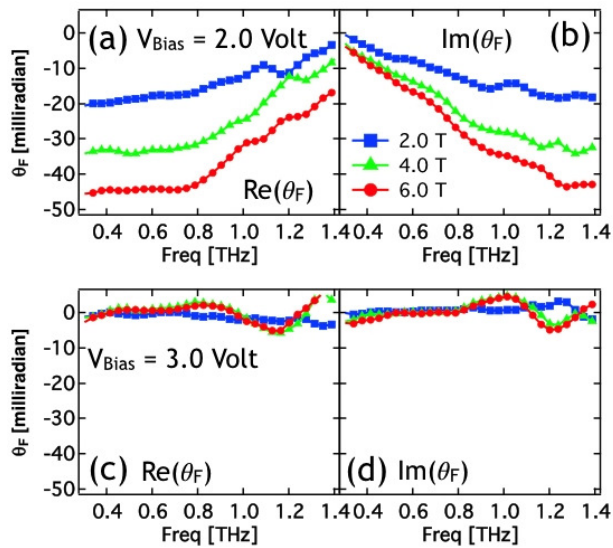

FIG. S3. (Color online) Real and imaginary parts of the Faraday angle of a 32 QL $Bi_2Se_3$ film without the $MoO_3$ protection layer. (a-b) Faraday rotation measured at 2.0 V which show the usual behavior consistent with the previous measurements [S6]. (d-e) The measured Faraday rotation at 3.0 V is almost zero. We observed a large transient current ($> 1$ $\mu A$) at around 2.5 Volt, while changing the bias voltage from 2.0 V to 3.0 V. This indicates the chemical reaction between the ionic gel and unprotected $Bi_2Se_3$ film, that permanently damage the film. The $MoO_3$ capping layer provide protection against chemical reactions and allows us to apply higher voltage ($> 5.0$ V).

The effects can be seen in Fig. S3. In Figs. S3a and b, we show the real and imaginary parts of the Faraday angle of a 32 QL $Bi_2Se_3$ film not coated with the $MoO_3$ protection layer. At 2 V, the Faraday rotation shows the usual behavior consistent with the previous measurements [S6]. However on raising the bias to 3 V we observed an extra large transient current ($> 1$ $\mu A$) around 2.5 V. Upon subsequent measurement the Faraday angle fell to effectively zero as shown in Figs. S3c and d. This indicates a chemical reaction occurred between the ionic liquid and the unprotected $Bi_2Se_3$ film that permanently damaged the film. The $MoO_3$ capping layer provides protection against chemical reactions and allows us to routinely apply higher voltage ($> 5.0$ V).

* sspmm4@iacs.res.in